\documentclass{iauc}
\usepackage{epsf}
\usepackage{psfig}
\usepackage{graphics}

  \checkfont{eurm10}
  \iffontfound
    \IfFileExists{upmath.sty}
      {\typeout{^^JFound AMS Euler Roman fonts on the system,
                   using the 'upmath' package.^^J}%
       \usepackage{upmath}}
      {\typeout{^^JFound AMS Euler Roman fonts on the system, but you
                   dont seem to have the}%
       \typeout{'upmath' package installed. iaus.cls can take advantage
                 of these fonts,^^Jif you use 'upmath' package.^^J}%
      }
  \else
  \fi

  \checkfont{msam10}
  \iffontfound
    \IfFileExists{amssymb.sty}
      {\typeout{^^JFound AMS Symbol fonts on the system, using the
                'amssymb' package.^^J}%
       \usepackage{amssymb}%

      }{}
  \fi

  \IfFileExists{amsbsy.sty}
    {\typeout{^^JFound the 'amsbsy' package on the system, using it.^^J}%
     \usepackage{amsbsy}}
    {\providecommand\boldsymbol[1]{\mbox{\boldmath $##1$}}}

\newsavebox{\astrutbox}
\sbox{\astrutbox}{\rule[-5pt]{0pt}{20pt}}

\title[The Gaia astrometric and photometric survey]
      {Microarcsecond astrometry with Gaia: the solar system, the Galaxy and beyond}

\author[Coryn A.L.\ Bailer-Jones]%
{Coryn A.L.\ Bailer-Jones%
}

\affiliation{Max-Planck-Institut f\"ur Astronomie, K\"onigstuhl 17, 69117
Heidelberg, Germany\\ email: calj@mpia-hd.mpg.de}

\pubyear{2004}
\volume{196}
\pagerange{1--8}
\date{?? and in revised form ??}
\jname{Transit of Venus: New Views of the Solar System and Galaxy}
\editors{D.W. Kurtz \& G.E. Bromage, eds.}

\def\uas{$\mu$as}
\def\deg{$^\circ$}
\def\kms{kms$^{-1}$}

\def\teff{${\rm T}_{\rm eff}$}

\def\met{[M/H]}

\begin{document}

\maketitle

\begin{abstract}
Gaia is an all sky, high precision astrometric and photometric
satellite of the European Space Agency (ESA) due for launch in
2010. Its primary mission is to study the composition, formation and
evolution of our Galaxy.  Over the course of its five year mission,
Gaia will measure parallaxes and proper motions of every object in the
sky brighter than visual magnitude 20, amounting to a billion stars,
galaxies, quasars and solar system objects. It will achieve an
astrometric accuracy of 10\,$\mu$as at V=15 -- corresponding to a
distance accuracy of 1\% at 1\,kpc -- and 200\,$\mu$as at V=20. With
Gaia, tens of millions of stars will have their distances measured to
a few percent or better. This is an improvement over Hipparcos by
several orders of magnitude in the number of objects, accuracy and
limiting magnitude.  Gaia will also be equipped with a radial velocity
spectrograph, thus providing six-dimensional phase space information
for sources brighter than V\,$\sim$\,17.  To characterize the objects (which
are detected in real time, thus dispensing with the need for an input
catalogue), each object is observed in 15 medium and broad photometric
bands with an onboard CCD camera.  With these capabilities, Gaia will
make significant advances in a wide range of astrophysical topics. In
addition to producing a detailed kinematical map of stellar
populations across our Galaxy, Gaia will also study stellar structure
and evolution, discover and characterise thousands of exoplanetary
systems (extending down to about ten Earth masses for the nearest
systems) and make accurate tests of General Relativity on large
scales, to mention just some areas.  I give an overview of the
mission, its operating principles and its expected scientific
contributions.  For the latter I provide a quick look in five areas on
increasing scale size in the universe: the solar system, exosolar
planets, stellar clusters and associations, Galactic structure and
extragalactic astronomy.

\end{abstract}

\firstsection 

\section{Introduction}

Distance measurement has been historically one of the most fundamental
challenges in astronomy. To measure cosmic distances from the Earth's
surface we must generally rely on parallaxes, i.e.\ the apparent change in
position of an object relative to some other object (or reference
frame) brought about by a known displacement of the observer.  Astronomical
distance measurement is hard because these displacements (e.g.\ the
motion of the Earth around the Sun) are small compared to the
distances we want to measure (e.g.\ the distance to the Galactic
centre), that is we need to measure very small angular changes in
position.\footnote{This problem has been captured more elegantly by
Douglas Adams (1978): ``Space is big. Really big. You won't
believe how vastly hugely mindbogglingly big it is. I mean you may
think it's a long way down the road to the chemist, but that's just
peanuts compared to space.''}  The lack of accurate distances
has been, and continues to be, one of the most significant limitations
in studying the universe.

The topic of this conference -- the transit of Venus across the Sun --
is central to this issue, because the attempts to observe it were a
milestone in observational astronomy and in our potential to measure
distances.  Following first Halley's and then Deslise's outline of a
method to measure the solar parallax (and hence the Astronomical Unit,
AU) from timing a Venus transit, numerous expeditions were mounted to
observe the four transits occurring during the 18th and 19th
centuries. While the degree of accuracy and consistency of these
methods in determining the AU were not as high as hoped or expected
(for reasons discussed elsewhere in this volume), they nonetheless
made a vital contribution. The timing of transits, for example, was an
ingenious addition to our otherwise limited way of measuring cosmic
distances.

Now, some 370 years after the first transit observations of Mercury
and Venus in the 1630s, distance measurement in the universe remains a
fundamental issue in astronomy. It is vital for understanding the
structure and evolution of stars, the formation and composition of our
Galaxy and ultimately for tracing the origin of the universe. Almost
all aspects of astrophysics rely to some degree on accurate distances,
and this ultimately relies on astrometry: the measurement of positions
over time and derivations of parallaxes and proper motions from them.

The Gaia mission will mark a significant step forward in
astrometry. Following in the wake of the very successful Hipparcos
mission (ESA 1997), Gaia will extend Hipparcos' capabilities by
several orders of magnitudes and through this make significant
breakthroughs in numerous astrophysical topics. Table~\ref{comparison}
gives a brief comparison of the main capabilities of Gaia compared to
Hipparcos.

In this contribution I summarise the main aspects of the satellite,
its mission and its observational principles, and provide a sample of
expected science contributions. For further information the reader is
referred to the concept study report (ESA 2001) and its summary
(Perryman 2001). I will say little about the radial velocity
spectrograph on board Gaia: for this see the contribution in this
volume by Mark Cropper.

\begin{table}
\begin{center}
\begin{tabular}{lll}
{\bf Quantity}           & {\bf Hipparcos}  & {\bf Gaia}  \\ 
 & & \\
Magnitude limit          & V=12.4           & G=20 \\
Completeness limit        & V=7.3--9.0       & G=20 \\
No.\ of sources          & 120\,000         & 26 million to G=15 \\
                         &                  & 250 million to G=18 \\
                         &                  & 1000 million to G=20 \\
No.\ of quasars          & none             & 0.5--1 million \\
No.\ of galaxies         & none             & 1--10 million \\
Target selection         & input catalogue~~~~~~~  & onboard; magnitude limited\\
Astrometric accuracy     & $\sim$1000\,\uas & 2--3\,\uas\ at G$<$10 \\
                         &                  & 5--15\,\uas\ at G=15\\
                         &                  & 40--200\,\uas\ at G=20\\
Broad band photometry    & 2 (B$_{\rm T}$,V$_{\rm T}$)   & 4--5 bands \\
Medium band photometry~~~~~~~   & none             & 10--12 bands \\
Spectroscopy             & none             & R=11\,500 (848--874\,nm) \\
Radial velocities        & none             & $\sigma$=1--10\,\kms\ to G=17--18 \\
\end{tabular}
\caption{Comparison of the capabilities of Gaia with its predecessor
Hipparcos.}
\label{comparison}
\end{center}
\end{table}

\section{Global astrometry}

\begin{figure}
\centerline{
\epsfxsize 9cm\epsfbox{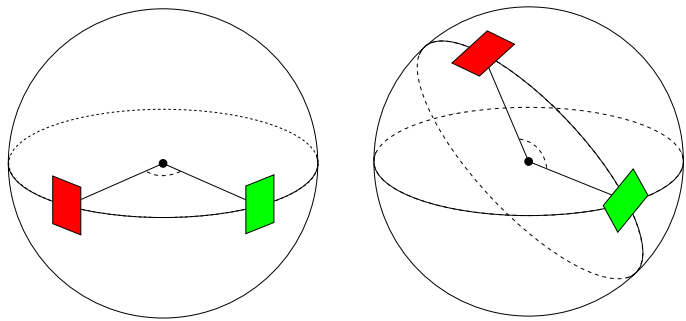}
}
\caption{The key to global astrometry is to measure positions
simultaneously in two fields separated by a large, fixed basic
angle. By repeating this for a given field at different orientations
(to give different reference stars) we can, in principle, determine
absolute parallaxes.}
\label{basic_angle_both}
\end{figure}

Astrometry is the practice of accurately measuring the positions of
objects on the celestial sphere. At any instant the position of a
celestial object is given by two co-ordinates, e.g.\ Right Ascension
and Declination. By measuring positions repeatedly over time, we can
measure parallax (due to the cyclic motion of the observing platform,
e.g.\ the Earth about the Sun) and proper motions (linear motions
through space in some reference frame). Combined with the radial
velocity, these yield six astrometric parameters comprising three
position co-ordinates and their first order time derivatives (three
velocity co-ordinates).
\footnote{Deviations from this six parameter model are important for
unresolved binary stars, which will show imposed Keplerian motions.
Moreover, higher order time derivatives (e.g.\ accelerations) can in
principle also be measured from an astrometric time series, and in
fact will be important with Gaia for some bright, nearby stars (the
{\it perspective acceleration}).}

Astrometry performed from the ground is done so over a narrow field.
That is, stellar positions can only be measured accurately relative to
nearby stars, typically within the field-of-view of the
telescope. Thus the derived parallaxes (in particular) are only
relative: all stars in the same direction share a common parallactic
effect due to the motion of the Earth about the Sun. To perform {\it
global astrometry} over the entire celestial sphere we must measure
the positions of stars relative to other stars separated by a large
angle on the sky, such that they have a different parallactic effect.
Repeating this (for a given field) for many different positions angles
(`reference fields'), and then measuring the positions of star in
{\it those} reference fields with respect to yet other stars separated
by large angles, we can eventually build up an entire grid of relative
position measurements over the celestial sphere (Fig.\
\ref{basic_angle_both}).  Note that we only accurately measure
positions in one dimension, i.e.\ along the great circle arc
connecting the two fields. Two dimensional positions are obtained from
the fact that we measure along great circles inclined at a range of
orientations.  We know that there are $2\pi$ radians in any great
circle so we can use this `closure condition' to derive absolute
positions of the objects. The choice of axes (e.g.\ RA and Dec) and
zero point (first point of Aries) are then a matter of convention. In
practice, the reduction of the data involves a large iterative
solution to simultaneously derive the five astrometric parameters of a
large number of stars.

The key to global astrometry is therefore to measure simultaneously 
the positions of stars in two widely separated fields-of-view and to
observe stars spread over the entire celestial sphere several times every
year for several years. (The last condition is imposed by the need
to lift the degeneracy between parallax and proper motion.)  To achieve
this in practice we must observe from space (also to overcome complex
refraction problems in the Earth's atmosphere).

This principle was used by the Hipparcos satellite, the first (and so
far only) mission to perform global astrometry (Perryman et al.\ 1989,
Kovalevsky 1995).

\begin{figure}
\centerline{
\epsfxsize 9cm\epsfbox{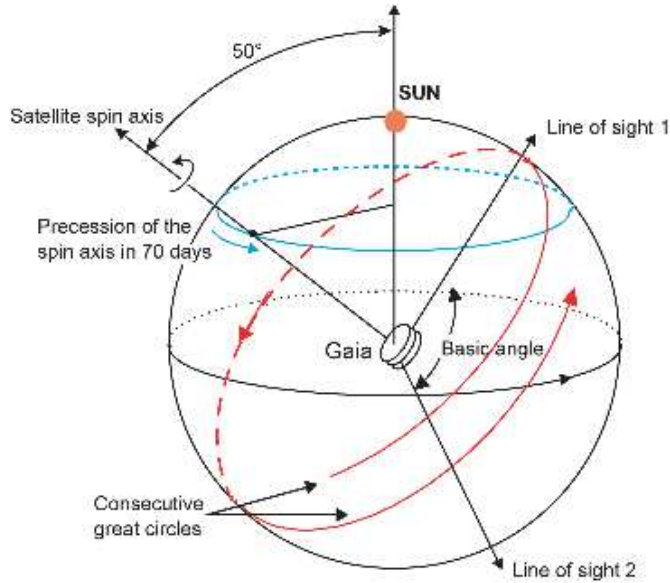}
}
\caption{The Gaia measurement principle and scanning law. Gaia
simultaneously observes in two viewing direction separated by a
constant {\em basic angle}. Gaia continuously rotates about an axis
perpendicular to these two viewing directions with a period of 6
hours.  This axis precesses in such a way that it maintains a constant
angle with respect to the Sun of 50\deg. The precession period is 70
days, so that every 6 hours Gaia traces a quasi-great circle across
the sky. From the combination of these two motions Gaia can observe
the whole sky. \copyright~ESA}
\label{scanning}
\end{figure}

\begin{figure}
\centerline{
\psfig{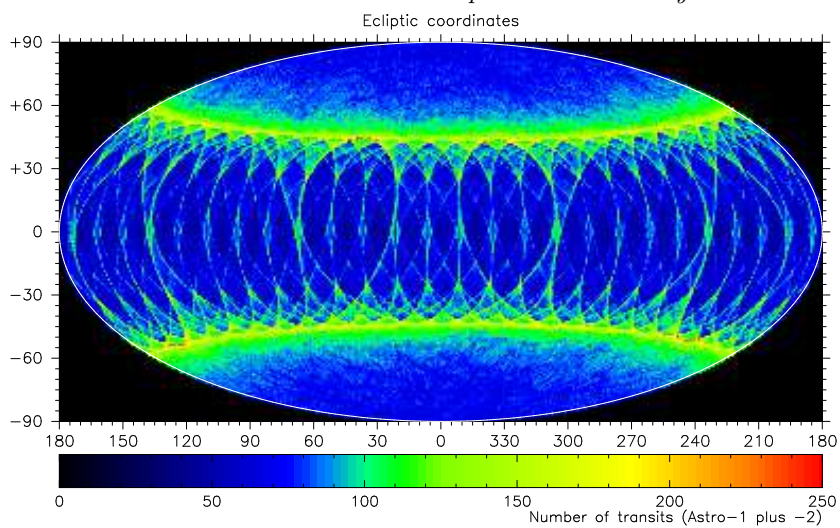}
}
\caption{Sky coverage by Gaia. The two-axis motion of Gaia (rotation
and precession) shown in Fig.~\ref{scanning} results in a non-uniform
sky coverage. The coverage depends on the specific scanning law
parameters, the size of the field-of-view and the duration of the
mission. This shows the number of observations at each point on the
sky for the nominal Gaia mission (5 years), plotted in ecliptic
co-ordinates.  \copyright~J.\ de Bruijne / ESA}
\label{path}
\end{figure}

The {\it scanning law} describes how the satellite observes the sky,
i.e.\ how the two fields shown in Fig.~\ref{basic_angle_both} move with
time.  With Gaia, it essentially consists of a three-axis
motion. First, Gaia rotates about its spin axis with a period of six
hours.  The two viewing directions lie in a plane perpendicular to
this axis (Fig.~\ref{scanning}).  To ensure it can observe the whole
sky, this spin axis simultaneously precesses (with a period of 70
days) such that it maintains a constant angle of 50\deg\ with respect
to the Sun (the {\it Sun aspect angle}, SAA -- the Sun is chosen for
reasons which are described below). As it rotates and precesses, Gaia
traces quasi great circles across the sky.  Finally, the satellite
orbits the Sun with a period of 1 year. The paths traced out by this
somewhat complex motion are shown in Fig.~\ref{path}.

\begin{figure}
\centerline{
\epsfxsize 9cm\epsfbox{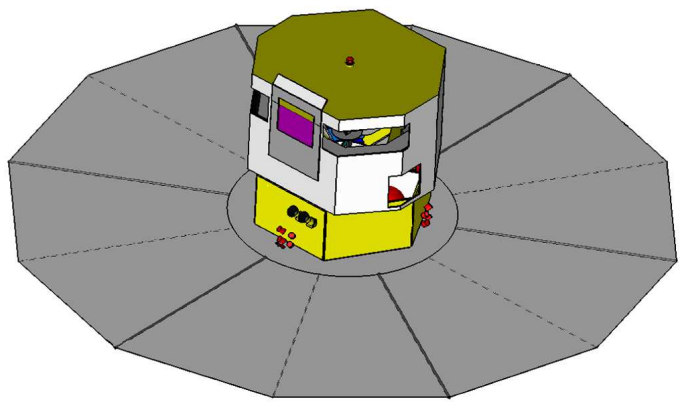}
}
\caption{The Gaia satellite and sun shield. The payload module is 3.1m
high and the sun shield has a diameter of 11m. Between the two entrance
slits for the two astrometric fields at the top is the astrometric
focal plane.  \copyright~Astrium}
\label{satellite}
\end{figure}

The angle between the two viewing directions -- the {\it basic angle}
-- is fixed. It is not necessary to know this angle exactly (it can be
derived as part of the reduction given the close condition) but it is
essential that it remain fixed, or at least that small variations of
it be measured. Specifically, to reach microarcsecond accuracy, the
basic angle must be constant to a few \uas\
over the six hour spin period. This places extremely stringent
requirements on the thermal and mechanical stability of the satellite.
For example, thermal gradients across the optical bench must be kept
below 25\,$\mu$K. Mechanical drags (e.g.\ from the Earth's tenuous
atmosphere) or thermal loads (e.g.\ from passing into the Earth's
shadow) could make achieving this almost impossible.  For this reason,
Gaia will be placed in orbit about the Earth--Sun L2 Lagrange point,
situated about 1.5 million km from the Earth (four times the
Earth--Moon distance) in the antisolar direction.\footnote{The L2
point itself in unstable so Gaia performs Lissajous orbits about
it. The orbit is such that Gaia does not pass into the Earth's
shadow during its five years of science operations.}  As the Earth and
Sun now lie in the same direction from Gaia, fixed in its rotating
reference frame, it precesses about this axis in order to avoid ever
observing the Earth or Sun. With SAA\,=\,50\deg\ Gaia never observes
closer than $90^{\circ} - 50^{\circ} = 40^{\circ}$ from the Earth/Sun.
Furthermore, with a flat sun shield deployed perpendicular to its spin
axis (Fig.~\ref{satellite}), this arrangement also ensures that the
solar flux on the spacecraft is constant as it precesses and orbits
the Sun.

\section{Gaia satellite}

\begin{figure}
\centerline{
\epsfxsize 9cm\epsfbox{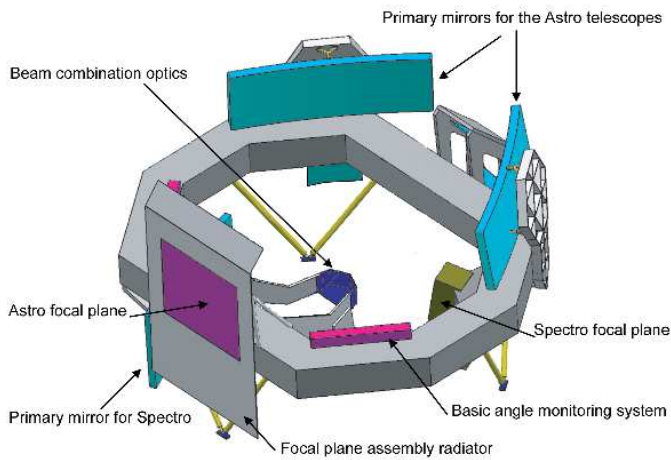}
}
\caption{The Gaia payload. The two astrometric primary mirrors are 1.4\,m\,$\times$\,0.5\,m in size. \copyright~Astrium}
\label{payload}
\end{figure}

The scientific payload of Gaia is built around a rigid optical bench
(Fig.~\ref{payload}). Onto this are mounted two astrometric telescopes
which image the two viewing directions onto a common focal plane
(together these form the {\em Astro} instrument). This focal plane
consists of a very large number of CCD detectors: there are some 170
arranged into a rectangular grid. Each CCD contains
4500$\times$\,1966 pixels with a pixel scale of
44\,mas\,$\times$\,133\,mas. As the satellite rotates, stars enter the
focal plane from the left and drift across it at a rate of
1\deg\,min$^{-1}$, crossing the focal plane in about 55\,s. The CCDs
are clocked in time-delayed integration (TDI) mode at the
same rate to track the stars. The reason for having a large number of
CCDs in the along-scan direction is to acquire a high signal-to-noise
ratio (SNR) in the stars' point spread function (PSF). This is the key
ingredient to precise centroiding of the PSF and hence precise angular
measurements in the along scan (quasi-great circle) direction.

Gaia detects objects in real time. There is no input catalogue telling
it approximately where the target objects are, as was the case with
Hipparcos.  With Gaia, anything with a point source magnitude of
G\,$<$\,20 is detected in real time (the G band is defined
below). This is done by the {\em star mapper} CCDs at the leading edge
of the focal plane: an initial strip of CCDs does the initial
detection, and another the subsequent confirmation and rejection of
cosmic rays etc. There is a separate strip of star mappers for each
field-of-view, masked such that each can only see one field-of-view.
The reason for these star mappers is two-fold.  First, there exists no
sufficiently accurate catalogue complete to G$=$20 over the whole sky
at the required spatial resolution which could act as an input
catalogue. Second, there is not enough bandwidth capacity in the Gaia
antenna to transmit the entire sky to the ground. To do this we would
require a data rate of around 1 Gb/s, whereas Gaia is limited to a few
Mb/s, i.e.\ more than a hundred times lower.\footnote{The reason for
this limited capacity is that Gaia must use an electronically
steerable array rather than a mechanically movable one, which would
permit a larger data rate. A mechanically movable antenna is
incompatible with the very severe requirements on mechanical stability
onboard the satellite, which is essential for high accuracy
astrometry.} Thus by using the star mapper detections, Gaia electronically
allocates CCD windows (of size 265\,mas$\times$\,1590\,mas) to stars
and transmits only these parts of the focal plane to the ground.

The CCDs in the main part of the astrometric focal plane are
unfiltered to maximise the number of photons collected. The resulting
photometric passband is determined by the CCD quantum efficiency curve
and the reflectivity of the optics and is named the G band. To
maximise sensitivity for the typical star (which is cool and/or
reddened) the mirrors are silver coated. The net passband therefore
lies between 400 and 1000\,nm.  At the high level of centroiding
accuracy required (ca.\ 1 part in 1000), chromatic effects in the
optics with this broad passband are significant. The position of the
centroid of the stellar PSF depends on the colour of the star and this
must therefore be corrected for. This is the main task of the {\em Broad Band
Photometer} (BBP), a set of four or five filters attached to the CCDs
at the trailing edge of the focal plane. (A different colour filter is
attached to each of the final strips of CCDs so that all
five colours are determined essentially simultaneously with the
astrometry.)

The spectroscopic instrument, {\em Spectro}, is a separate telescope
mounted on the same optical bench as Astro (see
Fig.~\ref{payload}). It is a 3-mirror telescope with a single viewing
direction. At the focal plane, part of the field is intercepted by a
dichroic mirror. This reflects the red part of the light into the {\em
Radial Velocity Spectrograph} (RVS), described in the article by Mark
Cropper in this volume. The blue transmitted light -- as well as the
white light in the other part of the field -- arrives at another CCD
focal plane, the {\em Medium Band Photometer} (MBP).  As with Astro,
this consists of an array of individual CCDs, although not as many and
with a larger pixel scale than in the case of Astro.  This CCD focal
plane operates in the same way as the astrometric CCDs, namely clocked
in TDI mode to track stars as the satellite rotates and also with an
independent set of star mappers. Now, however, a different medium band
filter is attached to each strip of CCDs. As a star traverses the
focal plane we therefore sample its spectral energy distribution in
some 10--12 passbands.  The photometric system and number of filters
is not finalised but it will cover most of the wavelength region
between 250 and 1000\,nm. The purpose of MBP is to classify the
objects observed and determine the physical parameters (temperature,
metallicity, surface gravity etc.)  for the stars. Techniques for
doing this are discussed by Bailer-Jones (2002, 2003).

\section{Astrometric accuracy}

The design of the Gaia satellite, in particular the size of the
mirrors and focal plane, its scanning law and mission duration, are
driven primarily by the astrometric accuracy we want to achieve.
These in turn are dictated by the science goals, discussed in detail in
the Gaia Concept and Technology Study Report (ESA 2000). Essentially, to be
able to answer the main questions about the three-dimensional
structure of the Galaxy and to establish the kinematics of tracer
stars to sufficiently large distances and with sufficient accuracy,
there are two main requirements. The first is to achieve an
end-of-mission astrometric accuracy of 10\,\uas\ at G=15. The second
is for the survey to extend to G=20 and still retain good accuracy.

To first order, the uncertainty in an astrometric parameter (position,
parallax or proper motion), $\sigma_a$, is given by $\sigma_a \sim
1/\sqrt{N}$, where $N$ is the number of photons collected (Lindegren
1978, ESA 2000).  Large $N$ is achieved with a large aperture, a large
field-of-view (so that a given star will be observed more often with a
given scanning law), a large focal plane (so that each star is
observed for longer during a transit) and a long mission life (so each
star is observed many times).

\begin{figure}
\centerline{
\epsfxsize 9cm\epsfbox{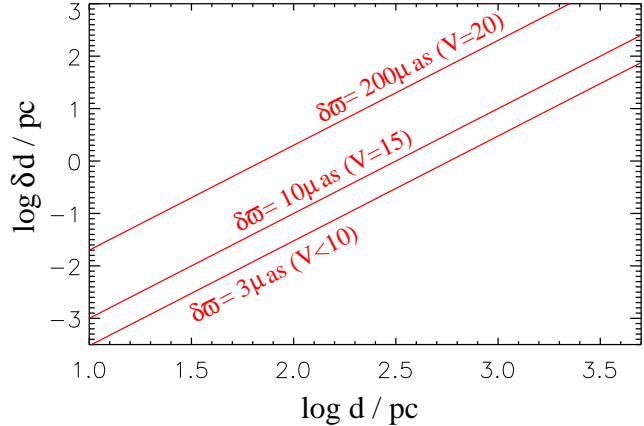}
}
\caption{The Gaia distance accuracy at three magnitudes. To first
order the uncertainty in the distance, $d$, is $\delta d \sim d^2
\delta \varpi$, where the parallax accuracy, $\delta \varpi$, depends
only on the apparent brightness (to first order) as $\delta \varpi
\sim 1/\sqrt{N}$ ($N$ is the number of photons collected).}
\label{astroacc2}
\end{figure}

Denoting the uncertainty in a parallax measurement with $\delta
\varpi$, the corresponding uncertainty in the distance, $d$, is
$\delta d \sim d^2 \delta \varpi$ (for $\delta \varpi \ll 1$). Thus
for a given parallax accuracy, the fractional distance accuracy
decreases linearly with increasing distance. For a given type of star
(fixed intrinsic luminosity), $N \sim 1/d^2$, and hence $\delta d \sim
d^3$.  Roughly speaking, given the five year mission of Gaia, the
astrometric and parallax accuracy are the same as each other (e.g.\
10\,\uas) and the same as the proper motion accuracy expressed per
year (10\,\uas\,yr$^{-1}$).  From this we can see how the distance
accuracy varies with distance (Fig.~\ref{astroacc2}).  Combining these
estimates with Galaxy models we can determine the number of objects
which obtain various accuracy levels (Table~\ref{accuracies}). We see
immediately the vast numbers of objects for which very accurate
distances are determined, e.g.\ some 20 million with distances better
than 1\% (which includes stars down to G=20 and with distances of up
to a few kpc).  In comparison, Hipparcos only obtained this distance
accuracy for stars brighter than V=10 which were nearer than 10 pc, a
mere handful of objects (168 to be precise).

\begin{table}
\begin{center}
\begin{tabular}{rlrrlr}
\multicolumn{3}{l}{\em distance accuracy} & \multicolumn{3}{l}{\em tangential velocity accuracy} \\
0.2\%  &  at d\,=\,200\,pc & at V\,=\,15~~~~~~~  &  0.1\,\kms  &  at d\,=\,2\,kpc  & at V\,=\,15  \\
1\%    &  at d\,=\,1\,kpc  & at V\,=\,15~~~~~~~  &    1\,\kms  &  at d\,=\,20\,kpc & at V\,=\,15  \\
20\%   &  at d\,=\,1\,kpc  & at V\,=\,20~~~~~~~  &    1\,\kms  &  at d\,=\,1\,kpc  & at V\,=\,20  \\
\multicolumn{3}{l}{\em accuracies achieved} & \multicolumn{3}{l}{\em accuracies achieved} \\
$<$\,0.1\%  &  \multicolumn{2}{l}{for 700\,000 stars}   & 0.5\,\kms & \multicolumn{2}{l}{for 44 million stars} \\
$<$\,1\%    &  \multicolumn{2}{l}{for 21 million stars} & 1\,\kms   & \multicolumn{2}{l}{for 85 million stars} \\
$<$\,10\%   &  \multicolumn{2}{l}{for 220 million stars}        & 5\,\kms   & \multicolumn{2}{l}{for 300 million stars} \\
\end{tabular}
\caption{Predictions (from ESA 2000) of the distance and tangential
velocity accuracy which Gaia will achieve. It may be useful to recall than
at a distance of 200\,pc, a velocity of 1\kms\ gives rise to a proper
motion of 1\,mas\,yr$^{-1}$.}
\label{accuracies}
\end{center}
\end{table}

\section{Scientific contributions}

A brief overview of Gaia's capabilities and likely contributions in
five scientific fields is now presented. For more information and
references see the Gaia Concept and Technology Study Report (ESA 2000)
and the summary by Perryman et al.\ (2001), from which much of this
section has been drawn. For more up-to-date information see the Gaia
website \verb+http://www.rssd.esa.int/gaia+

\subsection{Solar system}

Due to its high proper motion sensitivity, Gaia will detect numerous
minor planets in our solar system. In particular it is expected that 
it will detect some 10$^5$--10$^6$ new objects in the main asteroid belt,
compared to fewer than $10^5$ currently known.  The detailed
statistics of orbital elements so derived can be used to investigate
the formation of the solar system.
Knowing their distances, the apparent magnitudes of asteroids can be
used to derive their albedos. The medium and broad band photometry
from Gaia will be used to classify asteroids and thus derive their
chemical compositions. Furthermore, it is expected that during the
lifetime of Gaia some 100 close mutual encounters will be observed
and their open (hyperbolic) orbits traced: from this asteroid masses
can be determined. This would increase the number of directly (i.e.\
dynamically) determined asteroid masses from the present number of
about 20. Combined with size measurements from stellar occultations,
densities may be derived, thus providing insight into their chemical
composition and thus that of the primordial solar environment.

Because Gaia will be able to observe nearer to the Sun (within 40\deg)
and which much higher accuracy than is possible from the ground, it
will make important contributions toward the study of Near Earth
Objects (NEOs). Some of these asteroids, such as those of the Apollo
and Athen groups, cross the Earth's orbit and so are potentially at
risk of collision. As Gaia performs real-time onboard detection, it
will detect and observe such objects many times over the course of its
mission. It should be able to discover and determine accurate orbits
for all NEOs with diameters larger than a few hundred metres.

The Trojans are asteroids situated around the Lagrange L$_4$ and L$_5$
points of planets, where stable orbits can be maintained for hundreds
of millions of years. Planetary perturbations and mutual close
encounters mean that they can have orbits elongated around the L$_4$
and L$_5$ points of their `host' planet. Trojans are known for the
outer planets, but in the inner solar systems the only known ones are
two Martian Trojans.  It is still not known whether the Trojans were
captured or formed in situ, or even whether they have similar
compositions to the main belt asteroids.  Due to the large area it
covers, Gaia will undertake a systematic survey for the Trojans of all
planets, including, for the first time, for Venus.

Gaia will just about be sensitive enough to detect the very faint, red
population in the Edgeworth-Kuiper Belt (EKB).  The EKB, situated
around 40\,AU from the Sun, is a population of icy bodies which appear
to be a remnant of the formation of the solar system. Pluto is its
largest member.  An improved census of the EKB members, their orbits
and intrinsic properties will provide further insight into the
formation of planetary systems.  Although Gaia's sensitivity will not
permit detections of large numbers of these very faint, cold objects,
it is predicted to discover some 50 new objects. Of these, 5--10 are
expected to be binaries.

\subsection{Exosolar planets}

Through its high precision, multi-epoch astrometry Gaia will be able
to detect the positional `wobble' of a star due to the gravitational
force of a companion. Not only will this allow Gaia to detect hundreds
of millions of visually unresolved binary stars, but the sensitivity
is high enough to detect motions caused by planetary mass companions.
To date, most exoplanets have been detected by the radial
velocity method, i.e.\ the variable Doppler shift of spectral lines
induced by relative motions, although
a few have been detected by astrometry (see the article by F.\
Benedict in this volume) and via photometric detection of
transits.
The astrometric amplitude of the motion of a star of mass $M_{\rm s}$
due a planet of mass $M_{\rm p}$ orbiting at a distance $a$ is
$$ \alpha / '' = \left( \frac{M_{\mathrm p}} {M_{\mathrm
s}}\right) \left( \frac{a / {\mathrm AU}}{d / {\mathrm pc}}\right) $$ where $d$ is the
distance to the system from the Earth. For example, the 47 Ursa Majoris system is predicted
to give an amplitude of 360\,\uas.
The astrometric method is most sensitive to large or long period
orbits, which is the opposite of the radial velocity (RV) technique.
Therefore the techniques are complementary. But because the
astrometric signature is observed in a two-dimensional plane we can
determine the inclination of the Keplerian orbit. This allows a
determination of the mass of the planet without the $\sin i$ ambiguity
inherent to the RV technique (in which we only observe the projected
orbital velocities).\footnote{With both the astrometric and RV
techniques we can only determine the mass of the planet (or rather $m
\sin i$ for the RV method) if we assume a mass for the star: just
astrometry or RV alone does not allow us to determine the individual
masses of unresolved systems.  However, in many cases, the stellar
mass can be determined with reasonable accuracy from Gaia's
photometry.}

Gaia will be most useful for finding planets around stars with
V$<13$, with good capabilities out to about d\,=\,200\,pc.
This covers a vast number of stars -- around 100\,000, far more than
have been observed with current RV planet surveys --
over a wide range of spectral types. With a mission
of five years, Gaia will be able to determine orbits for planets
with periods of up to about 10 years. Extrapolating from current known
systems (and only a limited part of the orbital parameter space has
been explored), it is expected that Gaia will detect some 5000 new
planetary systems (compared to some 120 known as of mid 2004) and
determine accurate orbital elements for 1000--2000 of them.  From the
above equation, we see that for a given size of orbit and mass of
star, the minimum mass planet which Gaia can detect decreases
linearly with distance. For the nearest stars -- closer than 10\,pc --
Gaia will be able to detect planets with 3\% of the mass of Juipter,
or 10 Earth masses. 

This extensive exoplanetary survey will provide detailed
statistics on the types of planetary systems which stars of different
mass, metallicity, and age do and do not host. This will mark a major
contribution to understanding the formation and evolution of planetary
systems.

Gaia should also be able to detect some planets when they transit
across their host stars and thus cause a dimming of the integrated
light received by Gaia. Of course this only occurs for near edge-on
orbits (i.e.\ $i \simeq 90$\deg). As Jupiter has a diameter about ten
times smaller than the Sun, its transit would cause the Sun to dim by
1\%. To detect this with a signal-to-noise of 10 in the G band
requires a photometric precision of 1 millimag, easily achieved in
single epoch photometry for stars brighter than about G\,=\,13--14.
Despite the relatively poor time sampling for transits, Gaia
observes millions of such stars.  Predictions have been made that of
order 6000 planets around F--K type stars with semi-major axes in the
range 0--2\,AU will be detected this way.  Incidentally, and in
keeping with the theme of this conference, a transit of Venus across
the Sun (radius ratio of 1/115) observed from a large distance (not
from the Earth!) causes a maximum dimming of 0.08 millimag. This is at
the level at which Gaia photometry will probably be dominated by
systematic errors so is unlikely to be detectable even for the
brightest stars.

\subsection{Star formation and stellar clusters}

Most star formation occurs in cold, dense, dark molecular
clouds. As gas is converted into stars via gravitational collapse (and
quite possibly via the dissipation of supersonic turbulence), the cloud
emerges from being an embedded cluster to an optically visible young
star forming region or open cluster, such as the dense Orion
Nebula Cluster or the less dense low mass star forming region
Taurus--Auriga (with ages of a few million years).  Such clusters 
evolve dynamically: internal encounters will slowly evaporate the lower
mass members and interactions with the Galactic tidal field and
passing interstellar clouds will disrupt all but the densest clusters.
Surveys of stellar clusters show a mark drop off in numbers beyond a
few hundred million years in age. Thus is appears that almost all
clusters will disperse into the Galactic field population over a
timescale somewhat less than 1\,Gyr.

By studying the structure and content of clusters -- in particular the
initial mass function and binarity -- at different ages and abundances
we therefore study the recent star formation history of the Galaxy.

Gaia will make some of its most fundamental contributions in this
area. First, by measuring very accurate distances for individual stars
in clusters, it will allow an accurate determination of stellar
luminosities across a wide range of evolutionary phases, thus making
vital tests of theories of stellar structure and evolution.  For the
nearest clusters, such as the Hyades, Gaia will determine the IMF
(corrected for binarity) down into the brown dwarf regime and down to
a solar mass for cluster as far away as 3\,kpc.
By accurately determining the
distances to subgiants, the position of the Main Sequence turn off can
be determined to much higher accuracy than is presently possible,
resulting in a significant improvement in age determinations through
isochrone fitting.

Of the 1000 clusters currently known (mostly within 2\,kpc), distances
are only known to about half of them, of which many are still highly
uncertain, by factors of two or more. According to WEBDA there are
some 70 clusters within 500\,pc, of which 20 are closer than
200\,pc. For all of these, Gaia will determine distances to {\em
individual} members brighter than V\,=\,15 to better than 0.5\%.  This
corresponds to a depth accuracy of 2.5\,pc at 500\,pc or better for
nearer cluster and/or brighter stars.  These data will provide a three
dimensional map of clusters and their mass
distribution. Combined with accurate three dimensional velocities from
the radial velocities and proper motions, we will be able to study the
dynamical evolution of clusters across a range of characteristics in
much more detail than has previously been possible.  Recalling that
Gaia's distance accuracy, $\delta d$, depends on distance, $d$, as
$\delta d \sim d^2$ for a star of given apparent magnitude, then at a
distance of 100\,pc the depth uncertainty is only 0.1\% or 0.1\,pc.
For a given type of star, i.e.\ of given intrinsic luminosity, $\delta
d \sim d^3$. Thus out to 200\,pc, Gaia will measure the distances to
all giants and all dwarfs earlier than about K5--M0V to better than 1
part in 250.  In terms of kinematics, all these stars will have their
tangential velocities determined to better than 1\,\kms. Extending the
range to 500\,pc, then giant star kinematics are still measured with
this precision or better, as are dwarfs earlier than mid A types.
Armed with this information we can investigate mass segregation,
ejection and the the dispersion of clusters.

The power of Gaia comes in the vast number of stars it observes over
the whole sky with a well-defined selection function.  By searching
for clusterings in the multidimensional space formed by the 3D spatial
co-ordinates, the 3D velocity co-ordinates, as well as ``astrophysical
co-ordinates'' (\teff, \met\ etc.\ obtained from Gaia photometry) we
can use the Gaia archive to detect new clusters, associations and
moving groups out to several kpc. This will provide a much more
complete, systematic and reliable survey for stellar clusters than has
hitherto been possible.

\subsection{Galactic structure}

Gaia's combination of 6D phase space information and astrophysical
parameter determination make it a unique instrument for determining
the large scale structure of our Galaxy.  It is here that its faint
limiting magnitude, all-sky coverage and determination of stellar
properties (especially metallicity) are particularly important. We
can, for example, investigate the age--metallicity relation (using
long-lived K and M dwarfs) with a much larger sample than has
been possible to date. A dynamical determination of the mass of the disk and
its dark matter content will likewise be possible, as is a
determination of the Galactic rotation curves out to large
Galactocentric distances.

Within the Galactic disk, Gaia will provide an accurate mapping of
star forming regions and spiral arms. Interstellar extinction will
ultimately limit how far Gaia can see in the Galactic plane, but by
identifying and measuring parallaxes of very luminous Cepheid
variables, for instance, we can map out structure to large distances.

Moving to the halo, the Gaia database will allow us to search for
stellar clusterings in the Galactic halo which may be hallmarks of
merger events.  In addition to the spatial overdensity searches
generally carried out with existing star-count data sets, we can apply
multidimensional cluster search algorithms to look for intrinsic
patterns in a space--velocity--astrophysical parameter space. This
would be relevant for finding halo streams which may not be spatially
distinct but nonetheless share a common orbit and/or star formation
epoch. Even at 10\,kpc, Gaia can determine tangential velocities to
0.5\,\kms\ for stars as faint as G=15 (i.e.\ with absolute magnitudes
less than zero, such as OB main-sequence stars and K giants). In the
outer halo ($> 20$\,kpc from the Galactic Centre), most stars will
have relatively uncertain parallaxes (parallax errors of greater than
20\% for G\,$>$\,15).  However, photometric `parallaxes' may then provide
a reasonable distance estimate from which proper motions may still be
converted to tangential velocities of useful precision. Moreover, as
intervening foreground objects will have much larger (and accurate)
parallaxes, these foreground `contaminants' can be distinguished from
the background tracers giants which may then be used as kinematic and
density tracers in the halo.

\subsection{Extragalactic}

Beyond our Galaxy, the parallaxes of individual objects will be
negligible. By averaging over a population, useful results may be
obtained. For example, the Large Magellanic Cloud (LMC) has a parallax
of about 20\,\uas. But by averaging over a suitable set of members, a
geometric distance to the LMC with an accuracy of about 1\% should be
obtainable.

In studying local group galaxies, Gaia proper motions will permit a
better distinction between stars in our Galaxy and those in external
galaxies. With a reliable selection, mean distances and motions of
local group galaxies can be determined. By determining 3D orbits
within the local group out to 1--2\,Mpc (which includes some 20
galaxies), it may be possible to probe fluctuations in the initial
density distribution of matter on cosmological scales.

An important aspect of Gaia is its direct optical observations of
quasars. From these, Gaia astrometry will be tied to a quasi-inertial
reference frame with an accuracy of better than 1\,\uas\,yr$^{-1}$.
Down to its limiting magnitude of G\,=\,20, it is predicted that Gaia
will detect around half a million quasars. These will be observed in
all of Gaia's 15 photometric bands at some 100 epochs from which the
classes of quasars and their variability may be studied.
Interestingly, Gaia will be sensitive to the motion of the solar
system barycentre about the Galactic Centre. This motion induces an
aberration in the position of distant objects such as quasars.  With
the Sun moving at 220\,\kms\ at a distance of 8.5\,kpc about the
Galactic Centre with a period of 250\,Myr, quasars will show an
apparent proper motion of about 4\,\uas\,yr$^{-1}$.

Gaia performs real-time onboard detection of everything with a point
source magnitude brighter than G\,=\,20. Thus transient events, in
particular supernovae, will be detectable. Based on the current known
supernovae rate, Gaia will catch about 50 supernovae {\it per day}. A
real-time classification and alerts system is envisaged for such
events, from which rapid follow up by other observatories will be possible.

Finally, because Gaia will measure distances accurately to very large
numbers of different types of stars, including RR Lyraes and Cepheids,
it will permit a much better calibration of primary distance
indicators than has hitherto been possible. This will provide a
geometrically calibrated distance scale which is independent of CMB or
cosmological models.

\section{Current status and data analysis}

At the time of writing (mid 2004) Gaia is planned for launch in
2011. Present efforts focus on several essential technology
developments on the industrial side, related, for example, to the
CCDs, SiC mirrors and FEEP thrusters. On the scientific side, the
major on-going effort is preparation for the data analysis. 

Gaia will produce data at a rate of about 1 MB/s for 5 years,
producing a total of some 100 TB raw data. This quantity of data is
not actually that large. The challenging issue with the Gaia data
processing it the complexity of the data. Due to the continuous
scanning of the satellite over a period of five years, the astrometric
data stream essentially consists of a strip of data focal plane data
2.6 million degrees in length. Each object typically appears 100 times
in this data strip. Thus the data reduction involves object matching
in the 7000 or so different quasi-great circle scans making up this
strip.  The relative one-dimensional positions of a few hundred
million stars must then be determined from which the five basic
astrometric co-ordinates of each of these stars are solved for in the
so-called {\em Global Iterative Solution} (GIS). This process must
simultaneously solve for the attitude of the satellite and the various
instrument calibration parameters. Numerous additional astrometric
reduction tasks much be included, such as solving for higher order
astrometric terms of binary star systems. Furthermore, the radial
velocity data and photometric data must also be included, often
simultaneously. For example, the colour of the star must be included
in the astrometric reduction, as must be the radial velocity for
nearby, high proper motion stars (as their parallaxes and proper
motions will not be constant with time).

The data analysis with Gaia is therefore a major challenge. A data
reduction prototype, GDAAS, has been running for a few years which is
studying these problems. Beyond this, efforts are also ongoing to
ensure that the resulting Gaia database can be properly exploited for
subsequent scientific work. These include experiments in data mining
and work on theoretical models against which Galactic structure and
kinematic data may be compared.

\begin{acknowledgments}
Gaia is a project of the European Space Agency (ESA) with guidance
from the scientific community organised into several working
groups under the auspices of the Gaia Science Team (GST).  It is
through the efforts of these people, ESA staff and our industrial
partners that Gaia has reached its present advanced stage of
development. Much of this article has drawn from the Gaia Concept and
Technology Study Report (ESA 2000), complied by the former Gaia
Science Advisory Group, from which more detailed information and
references can be obtained.
\end{acknowledgments}


\begin{thebibliography}{}

\bibitem[]{adams79} Adams D. 1978 The Hitchhikers guide to the
Galaxy, (radio series), BBC.

\bibitem[]{bailer-jones02} Bailer-Jones C.A.L. 2002 \textit{Ap\&SS}
\textbf{280}, 21--29.

\bibitem[]{bailer-jones03} Bailer-Jones C.A.L. 2003 In \textit{GAIA
spectroscopy, science and technology} (ed.\ U.\ Munari). ASP Conf.\
Ser.\ 298, pp. 199--208. Astron.\ Soc.\ Pacific.

\bibitem[]{esa97} ESA 1997 \textit{Hipparcos Venice '97} ESA-SP-402.

\bibitem[]{esa00} ESA 2000 \textit{GAIA: Composition, formation and
evolution of the Galaxy} ESA-SCI(2000)4

\bibitem[]{lindegren78} Lindegren L. 1978 In
\textit{Modern Astrometry} (ed.\ F.V.\ Prochazka, R.H.\ Tucker). IAU
Colloquium no.\ 48, pp.\ 197--217. Institute of Astronomy (University
Observatory), Vienna.

\bibitem[]{kovalevsky89} Kovalevsky J. 1995 \textit{Modern astrometry}
Springer, Heidelberg.

\bibitem[]{perryman89} Perryman M.A.C. \& Hassan H. et al. 1989 \textit{The
Hipparcos mission. Pre-launch status} ESA SP-111.

\bibitem[]{perryman01}
Perryman M.A.C. et al. 2001 \textit{A\&A} \textbf{369}, 339--363.

\end{thebibliography}
\end{document}